\documentclass[preprintnumbers,amsmath,amssymbm,prd]{revtex4}
\usepackage{epsfig}
\usepackage{graphicx}
\usepackage{amssymb}

\begin{document}
\title{Quasinormal modes and strong cosmic censorship in near-extremal Kerr-Newman-de
Sitter black-hole spacetimes}
\author{Shahar Hod}
\affiliation{The Ruppin Academic Center, Emeq Hefer 40250, Israel}
\affiliation{ }
\affiliation{The Hadassah Institute, Jerusalem 91010, Israel}
\date{\today}

\begin{abstract}
\ \ \ The quasinormal resonant modes of massless neutral fields in
near-extremal Kerr-Newman-de Sitter black-hole spacetimes are
calculated in the eikonal regime. It is explicitly proved that, in
the angular momentum regime ${\bar a}>
\sqrt{{{1-2{\bar\Lambda}}\over{4+{\bar\Lambda}/3}}}$, the black-hole
spacetimes are characterized by slowly decaying resonant modes which
are described by the compact formula
$\Im\omega(n)=\kappa_+\cdot(n+{1\over2})$ [here the physical
parameters $\{{\bar a},\kappa_+,{\bar\Lambda},n\}$ are respectively
the dimensionless angular momentum of the black hole, its
characteristic surface gravity, the dimensionless cosmological
constant of the spacetime, and the integer resonance parameter]. Our
results support the validity of the Penrose strong cosmic censorship
conjecture in these black-hole spacetimes.
\end{abstract}
\bigskip
\maketitle

\section{Introduction}

The dynamics of linearized matter and radiation fields in black-hole
spacetimes are characterized by a discrete family of complex
(decaying in time) oscillation modes. These exponentially damped
quasinormal resonances, which dominate the late-time relaxation
dynamics of composed black-hole-field systems, have attracted the
attention of physicists and mathematicians during the last five
decades (see \cite{Nol,Ber,Kon} for excellent reviews and detailed
lists of references).

The characteristic quasinormal resonance spectrum
$\{\omega(l,m;n)\}^{n=\infty}_{n=0}$ (here the dimensionless angular
parameters $\{l,m\}$ are respectively the spheroidal harmonic index
and the azimuthal harmonic index which characterize the linearized
perturbation modes of the composed black-hole-field systems) of an
asymptotically flat composed black-hole-field system is determined
by the linearized Einstein-matter field equations with the
physically motivated boundary conditions of purely ingoing waves at
the absorbing black-hole event horizon and purely outgoing waves at
spatial infinity \cite{Det,Noteds}. The fundamental black-hole-field
quasinormal resonant mode (the mode with the smallest value of
$\Im\omega$) determines the characteristic timescale
\begin{equation}\label{Eq1}
\tau_{\text{relax}} \equiv 1/\Im\omega(n=0)\
\end{equation}
for the decay (relaxation) of linearized perturbation fields in the
exterior regions of the curved black-hole spacetime.

Due to the mathematical complexity of the linearized Einstein-matter
field equations, the complex resonant spectra of most black-hole
spacetimes are not known in a closed (and compact) analytical form.
Instead, one is usually forced to use numerical techniques in order
to solve the linearized Einstein-matter field equations with the
appropriate physically motivated boundary conditions which determine
the composed black-hole-field quasinormal resonance spectra.

Near-extremal (rapidly-spinning) Kerr black holes are unique in this
respect. In particular, solving analytically the linearized
Einstein-matter field equations, it has been explicitly proved that
the fundamental (least damped) quasinormal resonant frequencies of
equatorial massless perturbation modes of near-extremal Kerr black
holes are characterized by the remarkably compact analytical
relation \cite{Hod1,Hod2,Als,Noteneq,Noteunit}:
\begin{equation}\label{Eq2}
\Im\omega(n)=2\pi T_{\text{BH}}\cdot(n+{1\over 2})\ \ \ ; \ \ \
n=0,1,2,...\  ,
\end{equation}
where $T_{\text{BH}}$ is the semi-classical Bekenstein-Hawking
temperature \cite{Bekt,Hawt} of the black-hole spacetime, which is
related to the classical surface gravity $\kappa_+$ of its outer
(event) horizon by the simple relation \cite{Bekt,Hawt}
\begin{equation}\label{Eq3}
T_{\text{BH}}={{\kappa_+}\over{2\pi}}\  .
\end{equation}
In a very interesting work \cite{DiEp}, it has recently been
demonstrated, using semi-analytical techniques, that near-extremal
Kerr-de Sitter black holes are also characterized by the simple
functional relation (\ref{Eq2}).

Interestingly, defining the dimensionless black-hole rotation
parameter
\begin{equation}\label{Eq4}
{\bar a}\equiv {{a}\over{r_+}}\
\end{equation}
(here $r_+$ is the radius of the black-hole outer (event) horizon
[see Eq. (\ref{Eq13}) below]), one finds \cite{Hod2} that scalar
perturbation modes of near-extremal charged and spinning Kerr-Newman
black-hole spacetimes with large enough angular momenta,
\begin{equation}\label{Eq5}
{\bar a}>{\bar a}_{\text{c}}(l)\  ,
\end{equation}
are also characterized by the compact analytical relation
(\ref{Eq2}) \cite{Notealn,DiGo,Hodep}. Here the critical (minimal)
black-hole rotation parameter ${\bar a}_{\text{c}}(l)$, above which
the quasinormal resonant modes of the near-extremal Kerr-Newman
black-hole spacetimes are characterized by the compact analytical
relation (\ref{Eq2}), depends on the angular harmonic index $l$ of
the linearized perturbation modes. In particular, it has been proved
analytically that \cite{Hod2}
\begin{equation}\label{Eq6}
{\bar a}^{\text{KN}}_{\text{c}}(l\gg1)={1\over2}\
\end{equation}
for near-extremal Kerr-Newman black holes in the eikonal
(geometric-optics) $l=m\gg1$ regime.

The main goal of the present paper is to study analytically the
quasinormal resonance spectra of massless neutral perturbation
fields in non-asymptotically flat charged and rotating
Kerr-Newman-de Sitter (KNdS) black-hole spacetimes. It is important
to note that black holes in asymptotically de Sitter spacetimes have
recently attracted much attention in the context of the intriguing
Penrose strong cosmic censorship (SCC) conjecture \cite{HawPen,Pen}.
In particular, it has been shown (see \cite{DiEp,CarCo,Jao,Hodnw}
and references therein) that the validity of the fundamental SCC
conjecture in asymptotically de Sitter spacetimes depends on the
existence of (at least) one black-hole-field perturbation mode with
the property
\begin{equation}\label{Eq7}
\Im\omega\leq{1\over2}\kappa_-\  ,
\end{equation}
where $\kappa_-$ is the surface gravity which characterizes the
inner (Cauchy) horizon of the black-hole spacetime
\cite{DiEp,CarCo,Jao,Hodnw}.

Using analytical techniques, we shall explicitly prove below that,
in the eikonal (geometric-optics) regime $l\gg1$, the quasinormal
resonant modes of the near-extremal KNdS black-hole spacetimes with
large enough angular momenta, ${\bar a}>{\bar
a}^{\text{KNdS}}_{\text{c}}({\bar\Lambda})$ (here
${\bar\Lambda}\equiv \Lambda r^2_+>0$ is the dimensionless
cosmological constant of the black-hole spacetime), are
characterized by the compact functional relation (\ref{Eq2}). In
particular, we shall determine the functional dependence ${\bar
a}^{\text{KNdS}}_{\text{c}}={\bar
a}^{\text{KNdS}}_{\text{c}}({\bar\Lambda};l\gg1)$ of the critical
black-hole rotation parameter above which the neutral large-$l$
fundamental perturbation modes of the near-extremal KNdS black holes
conform to the important inequality (\ref{Eq7}) \cite{Notescc}.

\section{Description of the system}

We analyze the quasinormal resonance spectra of massless neutral
fields which are linearly coupled to a non-asymptotically flat
Kerr-Newman-de Sitter black-hole spacetime of mass $M$, angular
momentum $J\equiv Ma$, electric charge $Q$, and cosmological
constant $\Lambda>0$. The line element of the curved black-hole
spacetime can be expressed in the form \cite{CarLam,Suz,Noteqa}
\begin{equation}\label{Eq8}
ds^2=-{{\Delta_r}\over{\rho^2}}\Big({{dt}\over{I}}-a^2\sin^2\theta{{dL_z}\over{I}}\Big)^2
+{{\Delta_{\theta}\sin^2\theta}\over{\rho^2}}\Big[{{adt}\over{I}}-(r^2+a^2){{dL_z}\over{I}}\Big]^2
+\rho^2\Big({{dr^2}\over{\Delta_r}}+{{d\theta^2}\over{\Delta_{\theta}}}\Big)\
,
\end{equation}
where the metric functions are given by \cite{CarLam,Suz}
\begin{equation}\label{Eq9}
\Delta_r\equiv r^2-2Mr+Q^2+a^2-{1\over3}\Lambda r^2(r^2+a^2)\  ,
\end{equation}
\begin{equation}\label{Eq10}
\Delta_{\theta}\equiv 1+{1\over3}\Lambda a^2\cos^2\theta\  ,
\end{equation}
\begin{equation}\label{Eq11}
\rho^2\equiv r^2+a^2\cos^2\theta\  ,
\end{equation}
and
\begin{equation}\label{Eq12}
I\equiv 1+{1\over3}\Lambda a^2\  .
\end{equation}

The zeroes of the radial metric function \cite{CarLam,Suz}
\begin{equation}\label{Eq13}
\Delta_r(r_*)=0\ \ \ \ \text{with}\ \ \ \ *\in\{-,+,\text{c}\}\
\end{equation}
determine the horizon radii which characterize the KNdS black-hole
spacetime (\ref{Eq8}). For generic KNdS black holes, there are four
distinct (non-degenerate) roots $r_0<0< r_-\leq r_+\leq
r_{\text{c}}$ to the characteristic equation (\ref{Eq13}), where
$r_-$ is the inner (Cauchy) horizon, $r_+$ is the outer (event)
horizon of the black hole, and $r_{\text{c}}$ is the radius of the
cosmological horizon.

In the next section we shall explicitly prove that the quasinormal
resonance spectra of near-extremal KNdS black holes can be
determined analytically. These black-hole spacetimes are
characterized by the dimensionless relation $(r_+-r_-)/r_+\ll1$, or
equivalently \cite{Notetag}
\begin{equation}\label{Eq14}
r^{-1}_+\Delta'_r(r_+)\ll1\  .
\end{equation}
It is worth noting that, taking cognizance of Eqs. (\ref{Eq9}),
(\ref{Eq13}), and (\ref{Eq14}), one finds that near-extremal
Kerr-Newman-de Sitter (NEKNdS) black-hole spacetimes are
characterized by the dimensionless relation ${\bar
a}^{\text{NEKNdS}}\simeq \sqrt{{{1-{\bar\Lambda}-{\bar
Q}^2}\over{1+{\bar\Lambda}/3}}}$, where the dimensionless black-hole
physical parameters $\{{\bar a},{\bar Q},{\bar\Lambda}\}$ stand
respectively for $\{a/r_+,Q/r_+,\Lambda r^2_+\}$. Note that, in the
${\bar\Lambda}\to0$ limit, this relation reduces to the familiar
dimensionless relation ${\bar a}^{\text{NEKN}}\simeq\sqrt{1-{\bar
Q}^2}$ which characterizes near-extremal Kerr-Newman (NEKN)
black-hole spacetimes.

\section{The quasinormal resonance spectra of near-extremal
Kerr-Newman-de Sitter black-hole spacetimes}

In the present section we shall use analytical techniques in order
to calculate the quasinormal resonance spectra of near-extremal
charged and rotating KNdS black-hole spacetimes. In particular, we
shall use the well established \cite{Goe,Mash,CarMir} relation
between the black-hole quasinormal resonant frequencies in the
eikonal large-$l$ regime and the unstable null circular geodesics
which characterize the corresponding black-hole spacetimes.

The quasinormal resonant modes which dominate the linearized
relaxation dynamics of neutral perturbation fields in asymptotically
de Sitter back-hole spacetimes are characterized by the physically
motivated boundary conditions of purely ingoing waves at the outer
(event) horizon of the black hole and purely outgoing waves at the
cosmological horizon of the spacetime \cite{Cham}:
\begin{equation}\label{Eq15}
\psi \sim
\begin{cases}
e^{-i\omega y} & \text{\ for\ \ \ } r\rightarrow r_+\ \
(y\rightarrow -\infty)\ ; \\ e^{i\omega y} & \text{ for\ \ \ }
r\rightarrow r_{\text{c}}\ \ \ (y\rightarrow \infty)\  ,
\end{cases}
\end{equation}
where the tortoise radial coordinate $y$ is defined by the
differential relation $dy=[(r^2+a^2)/\Delta_r]dr$.

As explicitly proved in \cite{Goe,Mash,CarMir}, in the eikonal
(geometric-optics) $l\gg1$ regime, the real parts of the black-hole
quasinormal resonant frequencies are directly related (proportional)
to the characteristic angular velocity $\Omega_{\text{c}}$ of null
particles which are trapped at the unstable null circular geodesic
of the black-hole spacetime. Likewise, the imaginary parts of the
complex resonant frequencies are given, in the eikonal large-$l$
regime, by the remarkably compact relation \cite{CarMir}
\begin{equation}\label{Eq16}
\Im\omega(n)=-i(n+{1\over2})\cdot|\gamma|\ \ \ \ ; \ \ \ \
n=0,1,2,...\ ,
\end{equation}
where the integer $n$ is the resonance parameter of the composed
black-hole-field perturbation mode, and \cite{CarMir}
\begin{equation}\label{Eq17}
\gamma=\sqrt{{{V_r''}\over{2{\dot t}^2}}}\
\end{equation}
is the Lyapunov exponent which characterizes the instability
timescale $\tau=\gamma^{-1}$ of the null circular orbit
\cite{Noteleak} (the dot symbol $\dot\ $ denotes a derivative with
respect to the proper time $\tau$). Here $V_r(r)$ [see Eq.
(\ref{Eq21}) below] is an effective radial potential which
determines the geodesic motions of test particles in the black-hole
spacetime. It is worth emphasizing again that, as shown in
\cite{CarMir}, Eq. (\ref{Eq17}) is valid in the asymptotic large-$l$
regime.

The geodesic motions of a test particle of proper mass $m$ in the
KNdS black-hole spacetime are characterized by three conserved
quantities $\{E,L_z,K\}$ which are respectively related to the
stationarity property of the spacetime geometry, to its axial
symmetry, and to the hidden symmetry of the black-hole geometry
\cite{CarLam,CarKN,Bar,Stuc}. In particular, the equatorial motions
of test particles in the KNdS black-hole spacetime are governed by
the geodesic equations \cite{CarLam,CarKN,Bar,Stuc}
\begin{equation}\label{Eq18}
{{dr}\over{d\lambda}}=\pm V^{1/2}_r(r)\  ,
\end{equation}
\begin{equation}\label{Eq19}
r^2{{dL_z}\over{d\lambda}}=-IP_{\theta}+{{aIP_r}\over{\Delta_r}}\ ,
\end{equation}
and
\begin{equation}\label{Eq20}
r^2{{dt}\over{md\lambda}}=-aIP_{\theta}+{{(r^2+a^2)IP_r}\over{\Delta_r}}\
,
\end{equation}
where
\begin{equation}\label{Eq21}
V_r(r)=r^{-4}[P^2_r-\Delta_r(m^2r^2+K)]\  ,
\end{equation}
\begin{equation}\label{Eq22}
P_r=IE(r^2+a^2)-aIL_z\  ,
\end{equation}
\begin{equation}\label{Eq23}
P_{\theta}=I(aE-L_z)\  ,
\end{equation}
and
\begin{equation}\label{Eq24}
K=I^2(aE-L_z)^2\  .
\end{equation}
Here the affine parameter $\lambda$ is related to the proper time
$\tau$ by the simple relation $\tau=m\lambda$
\cite{CarLam,CarKN,Bar,Stuc}.

The characteristic equatorial circular geodesics of the black-hole
spacetime are determined by the relations
\cite{CarLam,CarKN,Bar,Stuc}
\begin{equation}\label{Eq25}
V_r(r=r_{\text{c}})=0\ \ \ \ \text{and}\ \ \ \
V_r'(r=r_{\text{c}})=0\ ,
\end{equation}
which, taking cognizance of Eqs. (\ref{Eq9}), (\ref{Eq12}),
(\ref{Eq21}), (\ref{Eq22}), (\ref{Eq24}) and defining the angular
velocity parameter \cite{CarMir}
\begin{equation}\label{Eq26}
\Omega_{\text{c}}={{E}\over{L_z}}\  ,
\end{equation}
yield the coupled algebraic equations \cite{Notemm}
\begin{equation}\label{Eq27}
\big[(r^2_{\text{c}}+a^2)\Omega_{\text{c}}-a\big]^2=\Delta_r(r=r_{\text{c}})\cdot(a\Omega_{\text{c}}-1)^2\
\end{equation}
and
\begin{equation}\label{Eq28}
4r_{\text{c}}\Omega_{\text{c}}\cdot\big[(r^2_{\text{c}}+a^2)\Omega_{\text{c}}-a\big]=
\Delta'_r(r=r_{\text{c}})\cdot(a\Omega_{\text{c}}-1)^2\
\end{equation}
for the null circular geodesics of the KNdS black-hole spacetimes.

We shall now use analytical techniques in order to determine the
physical and mathematical properties which characterize the
equatorial null circular geodesics of near-extremal [see Eq.
(\ref{Eq14})] KNdS black-hole spacetimes. To this end, it proves
useful to define the dimensionless small physical parameters
\begin{equation}\label{Eq29}
x\equiv {{r_{\text{c}}-r_+}\over{r_+}}\ \ \ \ ; \ \ \ \ y\equiv
\Omega_{\text{c}}\cdot{{r^2_++a^2}\over{a}}-1\  .
\end{equation}
Substituting Eq. (\ref{Eq29}) into Eqs. (\ref{Eq27}) and
(\ref{Eq28}), and using the near-horizon expansions \cite{Notedh}
\begin{equation}\label{Eq30}
\Delta_r(r=r_{\text{c}})=r_+\Delta_r'(r=r_+)\cdot
x+{1\over2}r^2_+\Delta_r''(r=r_+)\cdot x^2 +O(x^3)\
\end{equation}
and
\begin{equation}\label{Eq31}
\Delta_r'(r=r_{\text{c}})=\Delta_r'(r=r_+)+r_+\Delta_r''(r=r_+)\cdot
x+O(x^2)\  ,
\end{equation}
one obtains the leading-order (with $x\ll1$ and $y\ll1$) equations
\begin{equation}\label{Eq32}
\Big({{a}\over{r^2_++a^2}}\Big)^2\cdot\big(2r^2_+x+a^2y\big)^2=\big[r_+\Delta_r'(r=r_+)\cdot
x+{1\over2}r^2_+\Delta_r''(r=r_+)\cdot
x^2\big]\cdot(a\Omega_{\text{c}}-1)^2\cdot[1+O(x,y)]\
\end{equation}
and
\begin{equation}\label{Eq33}
{{a}\over{r^2_++a^2}}\cdot\big(2r^2_+x+a^2y\big)=\big[\Delta_r'(r=r_+)+r_+\Delta_r''(r=r_+)\cdot
x\big]\cdot{{(a\Omega_{\text{c}}-1)^2}\over{4r_{\text{c}}\Omega_{\text{c}}}}\cdot[1+O(x,y)]\
\end{equation}
for the equatorial null circular geodesics of the near-extremal KNdS
black-hole spacetimes.

From the two coupled equations (\ref{Eq32}) and (\ref{Eq33}) one
finds the two dimensionless physical parameters
\begin{equation}\label{Eq34}
x={{\Delta_r'(r=r_+)}\over{r_+\Delta_r''(r=r_+)}}\cdot
\Bigg[{{1}\over{\sqrt{1-{{r^2_+}\over{8a^2}}\Delta_r''(r=r_+)}}}-1\Bigg]\
\end{equation}
and
\begin{equation}\label{Eq35}
y={{2r_+\Delta_r'(r=r_+)}\over{a^2\Delta_r''(r=r_+)}}\cdot
\Bigg[1-\sqrt{1-{{r^2_+}\over{8a^2}}\Delta_r''(r=r_+)}\Bigg]\  ,
\end{equation}
which characterize the near-horizon ($x\ll1$) null circular
geodesics of the near-extremal [$\Delta'_r(r_+)/r_+\ll1$, see Eq.
(\ref{Eq14})] KNdS black-hole spacetimes.

It is important to stress the fact that the physical requirement
$x,y\in\mathbb{R}$ implies that the analytically derived relations
(\ref{Eq34}) and (\ref{Eq35}) are valid for near-extremal charged
and spinning KNdS black holes in the dimensionless angular momentum
regime [see Eqs. (\ref{Eq4}), (\ref{Eq9}), (\ref{Eq34}), and
(\ref{Eq35})]
\begin{equation}\label{Eq36}
{\bar a}>{\bar
a}^{\text{KNdS}}_{\text{c}}({\bar\Lambda})=\sqrt{{{1-2{\bar\Lambda}}\over{4+{\bar\Lambda}/3}}}\
.
\end{equation}
As a consistency check, we note that the critical rotation parameter
${\bar a}^{\text{KNdS}}_{\text{c}}({\bar\Lambda})$ of the
Kerr-Newman-de Sitter black-hole spacetimes [see Eq. (\ref{Eq36})]
reduces, in the ${\bar\Lambda}\to0$ limit, to the critical rotation
parameter ${\bar a}^{\text{KN}}_{\text{c}}$ [see Eq. (\ref{Eq6})]
which characterizes the Kerr-Newman black-hole spacetimes. [It is
important to note that, for KNdS black-hole spacetimes, the value of
the maximally allowed dimensionless cosmological constant
${\bar\Lambda}_{\text{max}}$ is a monotonically increasing function
of the charge parameter ${\bar Q}$ from
${\bar\Lambda}_{\text{max}}=\sqrt{3}(2-\sqrt{3})$ for neutral
Kerr-de Sitter black holes \cite{Yosh} to
${\bar\Lambda}_{\text{max}}=1/2$ for maximally charged
Reissner-Nordstr\"om-de Sitter black holes. Thus, the critical
physical parameter ${\bar
a}^{\text{KNdS}}_{\text{c}}({\bar\Lambda})$ as given by Eq.
(\ref{Eq36}) is real].

From Eqs. (\ref{Eq21}), (\ref{Eq22}), (\ref{Eq24}), and
(\ref{Eq26}), one finds the relation
\begin{equation}\label{Eq37}
V_r={{I^2L^2_z}\over{r^4}}\Big\{\Big[\Omega_{\text{c}}(r^2+a^2)-a\Big]^2-\Delta_r(a\Omega_{\text{c}}-1)^2\Big\}\
.
\end{equation}
From Eqs. (\ref{Eq25}) and (\ref{Eq37}) one deduces the relations
\begin{equation}\label{Eq38}
\Big[\Omega_{\text{c}}(r^2+a^2)-a\Big]^2-\Delta_r(a\Omega_{\text{c}}-1)^2=
\Big\{\big[\Omega_{\text{c}}(r^2+a^2)-a\big]^2-\Delta_r(a\Omega_{\text{c}}-1)^2\Big\}'=0\
,
\end{equation}
which imply [see Eq. (\ref{Eq37})]
\begin{equation}\label{Eq39}
V_r''={{I^2L^2_z}\over{r^4}}\Big\{\big[\Omega_{\text{c}}(r^2+a^2)-a\big]^2-\Delta_r(a\Omega_{\text{c}}-1)^2\Big\}''\
.
\end{equation}
From Eq. (\ref{Eq39}) one finds
\begin{equation}\label{Eq40}
V_r''(r=r_{\text{c}})={{I^2L^2_z}\over{r^4_{\text{c}}}}
\Big\{8(\Omega_{\text{c}}r_{\text{c}})^2+4\Omega_{\text{c}}\big[\Omega_{\text{c}}(r^2_{\text{c}}+a^2)-a\big]-
\Delta_r''(r=r_{\text{c}})(a\Omega_{\text{c}}-1)^2\Big\}\ .
\end{equation}
Substituting Eq. (\ref{Eq29}) into Eq. (\ref{Eq41}), one obtains
\begin{equation}\label{Eq41}
V_r''(r=r_{\text{c}})=\Big({{IL_z}\over{r^2_++a^2}}\Big)^2\cdot\Big[{{8a^2}
\over{r^2_+}}-\Delta_r''(r=r_+)\Big]\cdot[1+O(x,y)]\
\end{equation}
for the near-horizon equatorial null circular geodesics of the
near-extremal KNdS black holes.

In addition, from Eqs. (\ref{Eq20}), (\ref{Eq22}), (\ref{Eq23}), and
(\ref{Eq26}), one finds the relation
\begin{equation}\label{Eq42}
\dot
t={{I^2L_z}\over{r^2}}\Big\{-a(a\Omega_{\text{c}}-1)+{{r^2+a^2}\over{\Delta_r}}
[\Omega_{\text{c}}(r^2_{\text{c}}+a^2)-a\big]\Big\}\  .
\end{equation}
Substituting Eq. (\ref{Eq27}) into Eq. (\ref{Eq42}), one obtains
\begin{equation}\label{Eq43}
\dot t(r=r_{\text{c}})={{I^2L_z(1-a\Omega_{\text{c}})}
\over{r^2_{\text{c}}}}\Big[-a+{{r^2_{\text{c}}+a^2}\over{\sqrt{\Delta_r(r=r_{\text{c}})}}}\Big]\
.
\end{equation}
Substituting Eq. (\ref{Eq30}) into Eq. (\ref{Eq43}), one finds the
near-horizon ($x\ll1$) relation
\begin{equation}\label{Eq44}
\dot
t^{-1}(r=r_{\text{c}})={{r^2_+}\over{I^2L_z(1-a\Omega_{\text{c}})(r^2_++a^2)}}
\sqrt{r_+\Delta_r'(r=r_+)\cdot
x+{1\over2}r^2_+\Delta_r''(r=r_+)\cdot x^2}\cdot[1+O(x)]\  .
\end{equation}
Substituting Eqs. (\ref{Eq29}) and (\ref{Eq34}) into Eq.
(\ref{Eq44}), one obtains
\begin{equation}\label{Eq45}
\dot t^{-1}(r=r_{\text{c}})={{r_+\Delta_r'(r=r_+)}
\over{I^2L_z\sqrt{16a^2-2r^2_+\Delta_r''}}}\cdot[1+O(x,y)]\  .
\end{equation}

Substituting Eqs. (\ref{Eq41}) and (\ref{Eq45}) into the
characteristic eikonal relation (\ref{Eq17}), and using the
expression \cite{Yosh}
\begin{equation}\label{Eq46}
\kappa_+={{\Delta_r'(r=r_+)}\over{2I(r^2_++a^2)}}\
\end{equation}
for the surface gravity of the outer black-hole horizon, one obtains
the simple equality [see Eq. (\ref{Eq14})]
\begin{equation}\label{Eq47}
\gamma=\kappa_+[1+O(r_+\kappa_+)]\ \ \ \ ; \ \ \ \ \
r_+\kappa_+\ll1\ ,
\end{equation}
which, taking cognizance of Eq. (\ref{Eq16}), yields the remarkably
compact functional relation
\begin{equation}\label{Eq48}
\Im\omega(n)=-i(n+{1\over2})\cdot\kappa_+\ \ \ \ ; \ \ \ \
n=0,1,2,...
\end{equation}
for the fundamental quasinormal resonant frequencies which
characterize the near-extremal ($r_+\kappa_+\ll1$) charged and
rotating KNdS black-hole spacetimes in the eikonal
(geometric-optics) regime.

\section{Black-hole quasinormal resonant frequencies and the Penrose
strong cosmic censorship conjecture}

The strong cosmic censorship conjecture, introduced by Penrose
almost five decades ago \cite{Pen}, asserts that, starting with
physically reasonable (spatially regular) initial conditions, the
dynamics of self-gravitating matter and radiation fields, which are
governed by the Einstein field equations, will always produce
globally hyperbolic spacetimes. If true, this physically important
conjecture guarantees that classical general relativity is a
deterministic theory.

It is well known that eternal black-hole spacetimes which possess
regular inner Cauchy horizons are not globally hyperbolic
\cite{Chan,M1,M2}. In particular, for eternal charged and rotating
black holes, the inner spacetime regions which are located beyond
the black-hole Cauchy horizons are characterized by the presence of
past directed null geodesics that terminate on the inner timelike
singularities of the eternal black-hole spacetimes
\cite{Chan,M1,M2}. Thus, the Einstein field equations may fail to
determine uniquely the future dynamics of physical observers who
fall into eternal black holes which contain regular inner Cauchy
horizons \cite{Chan,M1,M2}.

Intriguingly, however, as originally discussed by Penrose
\cite{Pen}, physically realistic (dynamically formed) black-hole
spacetimes are probably globally hyperbolic. In particular,
according to the mass-inflation scenario
\cite{M1,M2,M3,M4,M5,M6,M7,M8,NGE1,NGE2,NGE3}, remnant perturbation
fields that fall into dynamically formed black holes are infinitely
blue-shifted as they approach the inner Cauchy horizons. This
blue-shift mechanism may turn the pathological inner Cauchy horizons
of eternal black-hole spacetimes into singular hypersurfaces that
provide natural non-extendable boundaries to the inner spacetime
regions of dynamically formed black holes
\cite{M1,M2,M3,M4,M5,M6,M7,M8,NGE1,NGE2,NGE3}.

As discussed in \cite{CarCo,Jao} (see also \cite{DiEp,Hodnw} and
references therein), in asymptotically de Sitter black-hole
spacetimes, the final fate of the fundamental Penrose SCC conjecture
\cite{Pen} is determined by the dimensionless ratio between the
physical parameters $\Im\omega_0$ and $\kappa_-$, which respectively
characterize the {\it decay} rate of the linearized perturbation
modes in the exterior regions of the dynamically formed black-hole
spacetimes and the {\it amplification} rate of the infalling fields
as they approach the inner Cauchy horizons of the corresponding
black-hole spacetimes \cite{Notepan}.

In particular, the validity of the SCC conjecture in dynamically
formed non-asymptotically flat charged and spinning KNdS black-hole
spacetimes depends on the existence of (at least) one
black-hole-field perturbation mode which is characterized by the
property \cite{DiEp,CarCo,Hodnw,Jao}
\begin{equation}\label{Eq49}
{{\Im\omega}\over{\kappa_-}}\leq{1\over2}\  .
\end{equation}

Taking cognizance of the analytically derived functional relation
(\ref{Eq48}), and using the inequality $\kappa_+\leq\kappa_-$
\cite{M2} which characterizes the surface gravities of the
black-hole outer (event) and inner (Cauchy) horizons, one deduces
that the fundamental (least damped, $n=0$) neutral perturbation mode
of the KNdS black-hole spacetimes conforms to the dimensionless
relation (\ref{Eq49}). We therefore conclude that near-extremal
charged and rotating KNdS black-hole spacetimes in the dimensionless
physical regime (\ref{Eq36}) respect the fundamental Penrose strong
cosmic censorship conjecture \cite{Pen,Notents}.

\section{Summary}

The quasinormal resonance spectra which characterize the relaxation
dynamics of linearized neutral fields in non-asymptotically flat
near-extremal Kerr-Newman-de Sitter black-hole spacetimes have been
studied analytically in the eikonal (geometric-optics) regime. We
have proved that the characteristic relaxation rates of the composed
black-hole-field systems are determined by the surface gravities
[see Eqs. (\ref{Eq17}), (\ref{Eq46}), and (\ref{Eq47})] of the
black-hole outer (event) horizons.

Interestingly, we have revealed the existence of a critical
dimensionless black-hole rotation parameter ${\bar
a}^{\text{KNdS}}_{\text{c}}={\bar
a}^{\text{KNdS}}_{\text{c}}({\bar\Lambda})$, above which the eikonal
neutral perturbation modes of the near-extremal black holes become
long lived. In particular, using the well established relation
between the black-hole quasinormal resonant frequencies in the
eikonal large-$l$ regime and the physical properties of the unstable
null circular geodesics which characterize the corresponding
black-hole spacetimes \cite{Goe,Mash,CarMir}, we have derived the
remarkably compact analytical formula [see Eqs. (\ref{Eq36}) and
(\ref{Eq48})]
\begin{equation}\label{Eq50}
{{\Im\omega(n)}\over{\kappa_+}}=-i(n+{1\over2})\ \ \ \ \ \text{for}
\ \ \ \ \ {\bar a}>{\bar
a}^{\text{KNdS}}_{\text{c}}({\bar\Lambda})=\sqrt{{{1-2{\bar\Lambda}}\over{4+{\bar\Lambda}/3}}}\
\end{equation}
for the quasinormal resonant modes of the composed
near-extremal-Kerr-Newman-de-Sitter-black-hole-linearized-neutral-field
systems.

Finally, we have pointed out that the fundamental (least damped,
$n=0$) resonant mode of the near-extremal black-hole spacetimes
conforms to the inequality $\Im\omega/\kappa_-\leq1/2$ [see Eq.
(\ref{Eq49})] which is imposed by the fundamental SCC conjecture
\cite{Pen}. The compact analysis presented in this paper therefore
supports the validity of the Penrose strong cosmic censorship
conjecture in these non-asymptotically flat charged and spinning
Kerr-Newman-de Sitter black-hole spacetimes.

\bigskip


\noindent {\bf ACKNOWLEDGMENTS}
\bigskip

This research is supported by the Carmel Science Foundation. I would
like to thank Yael Oren, Arbel M. Ongo, Ayelet B. Lata, and Alona B.
Tea for helpful discussions.



\begin{thebibliography}{99}
\bibitem{Noteds} It is worth noting that linearized massless perturbation
fields in an asymptotically de Sitter black-hole spacetime are
characterized by purely outgoing waves at the cosmological horizon
of the black hole [see Eq. (\ref{Eq15}) below].
\bibitem{Noteneq} It is worth noting that, as explicitly proved in
\cite{Hod1,Hod2,Als}, the simple analytical relation (\ref{Eq2})
also characterizes part of the non-equatorial ($m<l$) quasinormal
resonant modes of near-extremal black holes.
\bibitem{Noteunit} We use natural units in which $G=c=\hbar=k_B=1$.
\bibitem{Notealn} It is worth noting that the result (\ref{Eq2}) for
scalar perturbation modes of near-extremal Kerr-Newman black holes
has been established analytically \cite{Hod2}. As for the
mathematically more complex case of coupled
gravitational-electromagnetic linearized perturbation modes of the
near-extremal charged and spinning Kerr-Newman black holes, there
are compelling numerical indications \cite{DiGo,Hodep} for the
validity of the relation (\ref{Eq2}).
\bibitem{Notescc} It is worth emphasizing again that, as
discussed in \cite{DiEp,CarCo,Jao,Hodnw}, the validity of the
Penrose strong cosmic censorship conjecture \cite{HawPen,Pen} in
asymptotically de Sitter spacetimes requires the existence of (at
least) one black-hole-field perturbation mode which respects the
inequality (\ref{Eq7}).
\bibitem{Noteqa} We shall assume that $Q\geq0$ and $a\geq0$ without loss
of generality.
\bibitem{Notetag} We use the $'$ symbol to denote a derivative
with respect to the radial coordinate $r$.
\bibitem{Noteleak} Note that the Lyapunov exponent $\gamma$ characterizes
the rate at which the null particles leak out of the unstable null
circular geodesic of the black-hole spacetime
\cite{Goe,Mash,CarMir}.
\bibitem{Notemm} Here we have taken $m\to0$ in Eq. (\ref{Eq21}) for
the null circular geodesics of the KNdS black-hole spacetimes.
\bibitem{Notedh} Here we have used the identity $\Delta_r(r=r_+)\equiv0$.
\bibitem{Notents} We would like to emphasize that this is not to say that the Penrose SCC conjecture \cite{Pen}
is violated by highly charged KNdS black-hole spacetimes outside the
regime (\ref{Eq36}). In particular, it is worth mentioning that it
has been explicitly proved in \cite{Hodnw} that charged matter
perturbation fields may also conform to the dimensionless inequality
(\ref{Eq49}) which, as discussed in \cite{CarCo,Jao} (see also
\cite{DiEp,Hodnw} and references therein), provides a necessary
condition for the validity of the SCC conjecture \cite{Pen} in
asymptotically de Sitter black-hole spacetimes.
\end{thebibliography}
\end{document}